\documentclass[preprint,superscriptaddress,tightenlines,amsmath,amssymb, aps,prm,]{revtex4-2}

\usepackage{graphicx}
\usepackage{dcolumn}
\usepackage{bm}
\usepackage[version=4]{mhchem}
\usepackage[colorlinks=false,linkcolor=black,citecolor=blue]{hyperref}
\usepackage{gensymb}
\usepackage{chemformula} 
\usepackage[T1]{fontenc} 
\usepackage{textgreek} 
\usepackage{amsmath}
\usepackage{amssymb}
\usepackage{color,soul}
\usepackage{textcomp}
\usepackage{multirow}
\usepackage{makecell}
\usepackage{bm}
\usepackage[normalem]{ulem}
\usepackage[colorlinks=false,linkcolor=black,citecolor=blue]{hyperref}

\definecolor{azure}{RGB}{51,153,255}

\begin{document}
\title{Structural and Optoelectronic Properties of Thin Film \ce{LaWN3}}

\author{Rebecca W. Smaha}
\email{Rebecca.Smaha@NREL.gov}
 \affiliation{National Renewable Energy Laboratory, Golden, Colorado 80401, USA} 
\author{John S. Mangum}
 \affiliation{National Renewable Energy Laboratory, Golden, Colorado 80401, USA} 
\author{Ian A. Leahy}
 \affiliation{National Renewable Energy Laboratory, Golden, Colorado 80401, USA} 
\author{Julian Calder}
 \affiliation{National Renewable Energy Laboratory, Golden, Colorado 80401, USA} 
\author{Matthew P. Hautzinger}
 \affiliation{National Renewable Energy Laboratory, Golden, Colorado 80401, USA} 
\author{Christopher P. Muzzillo}
 \affiliation{National Renewable Energy Laboratory, Golden, Colorado 80401, USA} 
 \author{Craig L. Perkins}
 \affiliation{National Renewable Energy Laboratory, Golden, Colorado 80401, USA} 
 \author{Kevin R. Talley}
 \affiliation{National Renewable Energy Laboratory, Golden, Colorado 80401, USA} 
  \author{Serena Eley}
 \affiliation{Colorado School of Mines, Golden, Colorado 80401, USA}
 \affiliation{University of Washington, Seattle, Washington 98195, USA}
 \author{Prashun Gorai}
 \affiliation{National Renewable Energy Laboratory, Golden, Colorado 80401, USA} 
\affiliation{Colorado School of Mines, Golden, Colorado 80401, USA} 
 \author{Sage R. Bauers}
 \affiliation{National Renewable Energy Laboratory, Golden, Colorado 80401, USA} 
\author{Andriy Zakutayev}
\email{Andriy.Zakutayev@NREL.gov}
 \affiliation{National Renewable Energy Laboratory, Golden, Colorado 80401, USA}


\begin{abstract}
Nitride perovskites are an emerging class of materials that have been predicted to display a range of interesting physics and functional properties, but they are highly under-explored due to the difficulty of synthesizing oxygen-free nitrides. \ce{LaWN3}, recently reported as the first oxygen-free nitride perovskite, exhibited polar symmetry and a large piezoelectric coefficient. However, the predicted ferroelectric switching was hindered by a large leakage current, which motivates better understanding of the electronic structure and optical properties of this material. In this article, we study the structure and optoelectronic properties of thin film \ce{LaWN3} in greater detail, employing combinatorial techniques to correlate these properties with cation stoichiometry. We report a two-step synthesis method that utilizes a more common radio frequency substrate bias instead of a nitrogen plasma source, yielding nanocrystalline films that are subsequently crystallized by ex-situ annealing. We investigate the crystal structure and composition of the combinatorial films produced by this method, finding polycrystalline La-rich films and highly textured W-rich films. The measured optical absorption onset (2.5 -- 3.5 eV) and temperature- and magnetic field-dependent resistivity (0.0055 -- 2 $\ohm$ cm for W-rich and La-rich \ce{LaWN3}, respectively) are consistent with semiconducting behavior and are highly sensitive to cation stoichiometry, which may be related to amorphous impurities: metallic W or \ce{WN_x} in W-rich samples and insulating \ce{La2O3} in La-rich samples. The fractional magnetoresistance is linear and small, consistent with defect scattering, and a W-rich sample has n-type carriers with high densities and low mobilities. We demonstrate a photoresponse in \ce{LaWN3}: the resistivity of a La-rich sample is enhanced by $\sim$28\% under illumination and at low temperature, likely due to a defect trapping mechanism. The physical properties of \ce{LaWN3} are highly sensitive to cation stoichiometry, like many oxide perovskites, which therefore calls for precise composition control to utilize the interesting properties observed in this nitride perovskite.
\end{abstract}

\maketitle

\section{\label{sec:Intro}Introduction}

The perovskite crystal structure, first described for oxides by Goldschmidt and Megaw \cite{Goldschmidt1926,Megaw1945}, is the largest and arguably the most flexible and useful family of crystalline materials. Perovskite and perovskite-derived materials display a staggering array of interesting ground states and functional properties, including high-temperature superconductivity in Ruddlesden-Popper layered cuprates, ionic and electrical conductivity in solid oxide fuel cells, ferroelectricity and dielectric behavior, and the photovoltaic effect in halide perovskites. 
Although many new ternary nitrides have recently been predicted, synthesized, and characterized \cite{Sun2019,Zakutayev2022}---including a number of anti-perovskite nitrides (\ce{\textit{M}3N\textit{E}})  \cite{Niewa2019}---phase-pure nitride perovskites have proven significantly less accessible, despite intriguing calculated physical properties. The first nitride perovskite, \ce{TaThN3}, was reported in 1995, although the amount of oxygen incorporation was not characterized \cite{Brese1995}.  It is predicted to be semiconducting, have a large Seebeck coefficient, and be a topological insulator \cite{Bannikov2007,Jung2018}, although little follow-up experimental work has been reported. In addition, many oxynitride perovskites have been synthesized with a wide variety of properties, from electrochemical activity to colossal magnetoresistance \cite{Bacher1988,Brese1995,Clarke2002,Yang2010,Fuertes2012,Black2016,Talley2019}, highlighting the possibility of uncovering interesting and potentially functional properties in the nearly unknown phase space of fully nitrided perovskites. 

Recent work has reignited the search for nitride perovskites, spurred not only by experimental work on oxynitrides but also by computational studies showing that rare earth-based compositions should be stable---or metastable, but synthetically accessible---and exhibit interesting physics, including magnetic and ferroelectric properties \cite{Sarmiento-Perez2015,Korbel2016,Flores-Livas2019,Grosso2023}. These studies identified in particular the \ce{\textit{RE}WN3} and \ce{\textit{RE}ReN3} (\textit{RE} = rare earth) families as the most likely to form a perovskite structure. Follow-up density functional theory (DFT) calculations on \ce{LaWN3} using hybrid functionals predicted that it should be a ferroelectric semiconductor with its ground-state structure as polar space group \textit{R}3\textit{c}, a large spontaneous polarization ($\sim$61 $\mu$C/cm$^2$), and a band gap of 0.9 -- 2 eV \cite{Sarmiento-Perez2015,Fang2017,Singh2018,Bandyopadhyay2020,Liu2020,MATSUISHI2022,Geng2023}. We previously successfully synthesized \ce{LaWN3} as an oxygen-free perovskite in thin film form using radio frequency (RF) co-sputtering with plasma-activated nitrogen \cite{Talley2021}. Recently, \ce{LaWN3} was synthesized in bulk using high pressure solid state metathesis to increase the chemical potential of nitrogen, requiring 5 GPa and 1300 $^\circ$C \cite{MATSUISHI2022}. Rietveld refinements could not distinguish between polar, rhombohedral \textit{R}3\textit{c} symmetry and tetragonal $I\overline{4}$ \cite{Talley2021} or cubic \textit{Pm}-3\textit{m} symmetry\cite{MATSUISHI2022}. However, the large piezoelectric response measured in thin film \ce{LaWN3} (effective piezoelectric
strain coefficient $d_{33,f}\approx$ 40 pm/V) confirms the calculated polar \textit{R}3\textit{c} symmetry and properties\cite{Talley2021}. Attempts to measure a ferroelectric response in these films were unsuccessful due to large leakage currents, implying that more in-depth study of the optical and electronic properties is necessary. 

Three additional rare earth nitride perovskites have recently been reported. Similar thin film deposition methods yielded \ce{CeWN3} and \ce{CeMoN3} \cite{Sherbondy2022}, finding that a competing fluorite phase was often present as an intermediate in the crystallization pathway from amorphous material to perovskite; heating the substrate during deposition or ex-situ annealing in flowing \ce{N2} were necessary to crystallize the perovskite phases of \ce{CeWN3} and \ce{CeMoN3}, respectively. \ce{CeMoN3} is antiferromagnetic below $T_N\approx 8$ K, while  \ce{CeWN3} is paramagnetic down to $T=2$ K; both have large, negative Weiss temperatures, suggesting strong magnetic frustration. In addition, bulk \ce{LaReN3} was synthesized using high-pressure techniques; interestingly, it required higher pressure (8 GPa and 1000--1200 $^\circ$C) to stabilize compared to \ce{LaWN3} \cite{Klos2021} even though calculations suggested it should be slightly more stable than \ce{LaWN3} \cite{Sarmiento-Perez2015}. \ce{LaReN3} was reported to have metallic transport and exhibit Pauli paramagnetism. 

In this article, we report on the structure and optoelectronic properties of thin film \ce{LaWN3}, employing combinatorial techniques to correlate these properties with stoichiometry. We report a two-step synthesis that utilizes RF substrate bias instead of an activated nitrogen plasma, as reported previously. 
We investigate the crystal structure and composition of the combinatorial films produced by this method, studying the trends in crystallinity and texture with composition. 
The measured optical absorption onset and resistivity are consistent with semiconducting behavior and are highly sensitive to cation stoichiometry, which may be related to amorphous impurities: metallic W or \ce{WN_x} in the W-rich samples and insulating \ce{La2O3} in the La-rich samples. The fractional magnetoresistance is linear and small, consistent with impurity scattering.  We demonstrate a photoresponse in \ce{LaWN3}; the resistivity is enhanced at low temperature, likely due to a trapping mechanism.

\section{\label{sec:Methods}Methods}

Experimental data used by this study have been analyzed using the COMBIgor software package \cite{Combigor} and are publicly available in the National Renewable Energy Laboratory (NREL) high-throughput experimental materials database at https://htem.nrel.gov \cite{HTEM}.

\subsection{\label{sec:Dep}Synthesis}
Thin films of \ce{LaWN3} were deposited using radio-frequency (RF) co-sputtering from elemental targets on 5.08 cm magnetrons in a vacuum sputtering chamber with a base pressure of approximately $2\times10^{-7}$ Torr. The powers used were 50 W or 54 W (W, 99.95 \%) and 73 W (La, 99.5 \%). Deposition occurred at a pressure of 4 mTorr under 5 sccm of Ar and 10 sccm of \ce{N2} (99.999\%) gases. The substrate was heated to approximately 700 $^\circ$C, and a cryogenic sheath was employed to trap adventitious oxygen or water during deposition. The source targets were presputtered for 90 or 120 minutes with the substrate shutter closed, followed by a 80 or 180 min deposition with a RF substrate bias (50 W). Immediately post deposition, the films were annealed for 30 min at 700 $^\circ$C (without sputtering or substrate bias) under 10 sccm of \ce{N2} at a pressure of 21.5 mTorr. Films were grown using the same conditions on two types of 5.08 $\times$ 5.08 cm substrates for different measurements:  p-type Si(100) (pSi) substrates with either native or 100nm \ce{SiO2} layers and sapphire (\ce{Al2O3}) substrates. All substrates had a thin film of W on the back as a refractory metal to improve substrate heating. This was removed from the sapphire substrates with \ce{H2O2} prior to optical measurements.

After removal from the sputtering chamber and initial structural and compositional characterization, the films were annealed in a ULVAC MILA-3000 rapid thermal annealing (RTA) furnace under flowing \ce{N2}. The films were heated over 12 min to 800 $^\circ$C and held at 800 $^\circ$C for 10 min, followed by rapid cooling.

\subsection{\label{sec:Struc}Composition and Structural Characterization}

Cation composition was measured with X-ray fluorescence (XRF) using a Fischer XUV XRF under vacuum ($\sim$1 Torr) using a model calibrated for pSi substrates (see Ref. \cite{Talley2021}).  For samples grown on sapphire substrates, the composition was assumed to be the same as pSi witness substrates. Oxygen and nitrogen contents were measured with Auger electron spectroscopy (AES) sputter depth profiling on a Physical Electronics 710 AES system.  A 10 kV, 10 nA primary beam was used.  Sputtering was performed with a 2 kV Ar$^+$ beam.  Nitrogen AES sensitivity factors were determined from binary metal nitride powders, and metal AES sensitivity factors were determined as described previously \cite{Davis1996} using one ungraded film grown on a pSi substrate for which the metal ratios were known from XRF.  Oxygen was quantified with an AES sensitivity factor from the analysis software package (Physical Electronics MultiPak v9.6.1.7).  Direct spectra were numerically differentiated and quantified within MultiPak.

Laboratory X-ray diffraction (XRD) patterns were collected with Cu K$_\alpha$ radiation on a Bruker D8 Discover diffractometer that allowed spatial mapping across the compositionally-graded thin films.
Select samples grown on sapphire substrates were also evaluated at beamline 11-ID-B at the Advanced Photon Source, Argonne National Laboratory. These synchrotron XRD data were collected in grazing incidence geometry with a wavelength of 0.1432 \AA.  LeBail fits were performed using the GSAS-II software suite \cite{GSAS}.

Samples for cross-sectional scanning electron microscopy (SEM) were prepared by mechanical cleaving immediately before loading them into the SEM vacuum chamber. Cross-sectional micrographs were acquired on a Hitachi S-4800 SEM operating at 3 kV accelerating voltage in secondary electron imaging mode.

\subsection{\label{sec:Prop}Optical, Transport, and Magnetic Characterization}
Transmission UV-vis optical spectroscopy spectra were collected on a Agilent Cary-6000i system with spot size approximately 3 mm$^2$ on films grown on sapphire substrates. Absorption coefficients were determined from these data together with thicknesses extracted from the SEM data. Carrier densities ($|n|$) were extracted by modeling the peak at low energy as the plasma frequency $\omega_p$; see Equation \ref{plasma} below.

Spectroscopic ellipsometry data were acquired at 65$^\circ$, 70$^\circ$, and 75$^\circ$ incident angles on a single row of a select annealed combinatorial sample (11 points per row) using a J.A. Woollam Co. M-2000 variable angle ellipsometer. The sample was grown on a crystalline pSi(100) substrate and was approximately 750-800 nm thick. CompleteEASE software (version 6.63) was used to model the data by fitting the real and imaginary parts of the dielectric function with a single layer model consisting of the \ce{LaWN_{3-y}} film. No substrate was necessary in the model due to the \ce{LaWN_{3-y}} film transmitting no photons through to the silicon substrate. The \ce{LaWN_{3-y}} film was modeled using two oscillators: (1) a parametric semiconductor oscillator (PSemi-Tri), which is an established model for accurately fitting the imaginary part of the dielectric function of crystalline semiconductor materials while maintaining Kramers-Kronig consistency\cite{Liu2017}, and (2) a Drude oscillator to model the sub-gap absorption below 2 eV.  

For electrical transport, $\sim$370 nm thick films were deposited as cloverleafs on a sapphire substrate. 10nm of Ti and 100nm of Au were deposited via electron-beam evaporation as electrical contacts for the Van der Pauw measurement. The sheet and Hall resistances ($R_S$ and $R_{xy}$) were measured as a function of temperature and applied magnetic field in a Quantum Design Physical Property Measurement System. All reported magnetoresistances (Hall resistances) are symmetrized (antisymmetrized) as a function of applied field.  
Temperature-dependent photoresistivity was additionally measured down to $T=100$ K using a Lake Shore Cryotronics Model 8425.
Samples were illuminated with a G2V Pico LED solar simulator with a 365 -- 1450 nm AM1.5G spectrum. We verify its intensity to be 23 mW/cm$^2$ using a reference solar cell.

Magnetic properties were measured on a \ce{LaWN_{3-y}} sample and a \ce{CeWN_{3-y}} sample via superconducting quantum interference device (SQUID) magnetometry in a Quantum Design Magnetic Properties Measurement System (MPMS3). The films were measured from 1.8--300 K under applied fields from -7 to +7 T. The measured \ce{LaWN_{3-y}} film was an approximately 5 $\times$ 5 mm piece of a combinatorial film grown on a pSi substrate with a thin layer of metallic W on the back. It was approximately 750 nm thick with composition La/(W+La) = 50.6\%, as measured by XRF. The measured \ce{CeWN_{3-y}} film was an approximately 5 $\times$ 5mm piece of a combinatorial film grown on a pSi substrate; it was approximately 150 -- 200 nm thick with cation composition Ce/(W+Ce) $\approx$ 51\% \cite{Sherbondy2022}. To isolate the signal of the films, bare substrates were also measured and subtracted. The resulting signal for \ce{CeWN_{3-y}} was scaled to match the approximate volume of the \ce{LaWN_{3-y}} film.

\subsection{Computations} \label{methods:calculations}
We used first-principles density functional theory (DFT) to perform structural relaxation of \ch{LaWN3}. All DFT calculations were performed with the Vienna Ab Initio Simulation Package (VASP)\cite{VASP1, VASP2} using projector augmented wave (PAW) pseudopotentials\cite{PAW2} to describe the core electrons. The wavefunctions were expanded as plane waves with an energy cutoff of 340 eV. The structural relaxations were performed within the generalized gradient approximation (GGA) of Perdew-Burke-Erzenhof (PBE) as the exchange correlation functional. A Hubbard on-site energy correction of $U$= 3 eV was applied to the $d$ orbitals of La and W, following the methodology in Refs. \citenum{Gorai2016_HubbardU} and \citenum{FERE}. For the relaxed structure, we calculated the electronic structure using the tetrahedron integration scheme on a dense $\Gamma$-centered {$8\times8\times8$} Monkhorst-Pack $k$-point grid. The density-of-states effective mass ($m^*_\mathrm{DOS}$) was determined from the DOS within the single parabolic band approximation, such that the parabolic band reproduces the same number of states as the DOS within a 100 meV energy window from the relevant band edges. The band effective mass ($m_{\mathrm{b}}^*$) was calculated within the parabolic and isotropic band approximation: $m_{\mathrm{b}}^*$ = $m_{\mathrm{DOS}}^*$ $N_{b}^{-2/3}$, where $N_b$ is the band degeneracy.\cite{qu2020} The static dielectric constant ($\epsilon_\infty$), used to determine the plasma frequency, was calculated using density functional perturbation theory. The underestimation of the band gap with GGA-PBE was remedied by computing band edge shifts from GW quasiparticle energies.\cite{gorai2021} We used DFT wave functions as initial wavefunctions for the GW calculations. GW eigenvalues were then iterated to self-consistency, removing the dependence on the single-particle Kohn-Sham energies. The input DFT wave functions were kept constant during the GW calculations, which allows the interpretation of the GW eigen energies in terms of energy shifts relative to the Kohn-Sham energies. The GW quasiparticle energies were calculated using on $\Gamma$-centered 8$\times$8$\times$8 $k$-point grid.

\section{\label{sec:Results}Results and Discussion}
\subsection{Growth and Composition}
Thin films of \ce{LaWN3} were grown via RF co-sputtering at high temperature ($\sim$700 $^\circ$C) on pSi and sapphire (\ce{Al2O3}) substrates to enable a range of property measurements. Combinatorial growth was performed targeting a narrow region of phase space around \ce{LaWN3}. Throughout this manuscript, we use \ce{LaWN_{3-y}} to refer to the combinatorial films, which generally have a small amount of N deficiency ($y\approx0.5$), as noted by Ref. \cite{Talley2021}. The growths yielded films with a composition gradient roughly $45\% < $ La/(W+La) $< 54\%$ (i.e., \ce{La_{0.9}W_{1.1}N_{3-y}} -- \ce{La_{1.08}W_{0.92}N_{3-y}}). 

\begin{figure}
\includegraphics[width=0.5\textwidth]{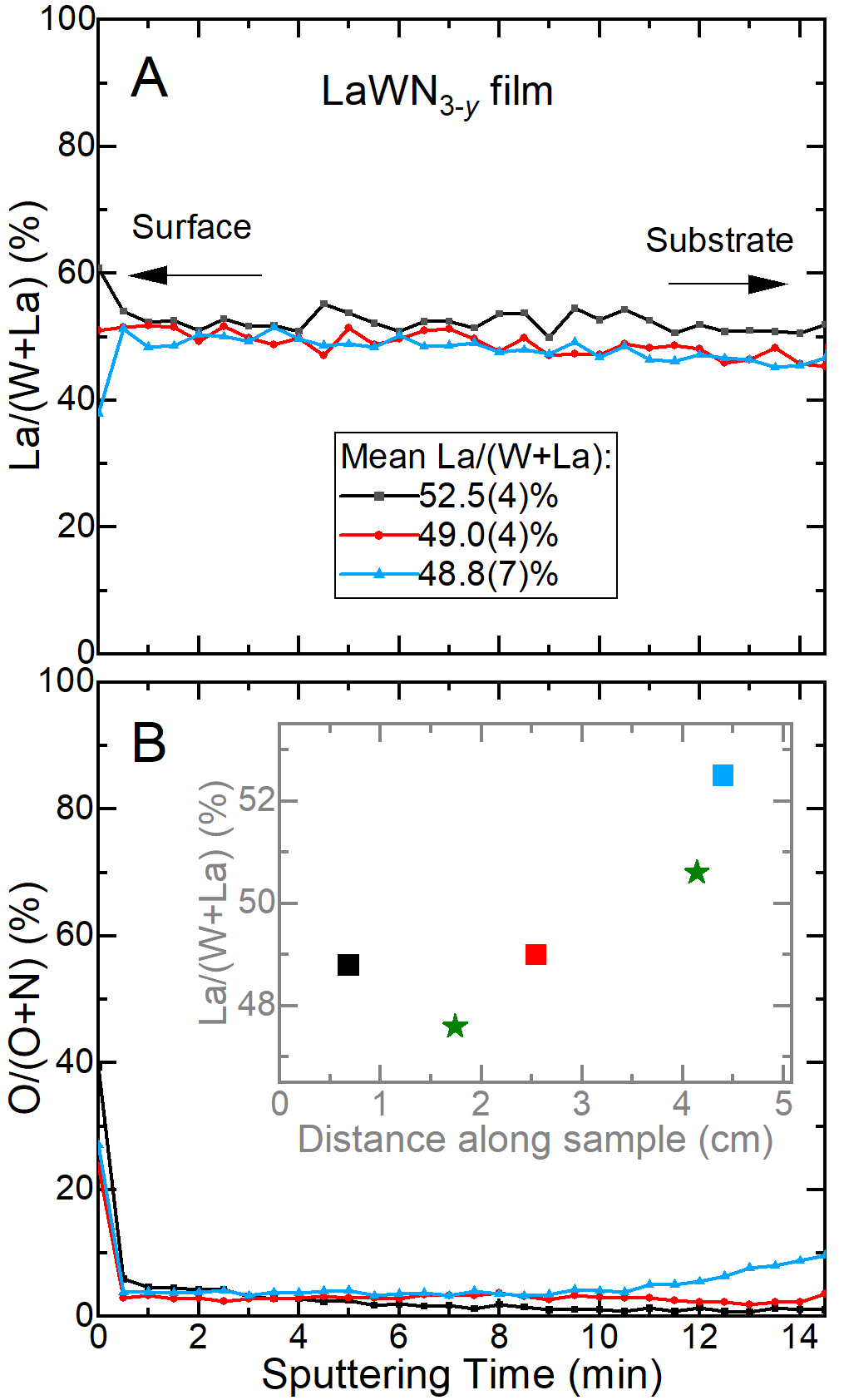}\\
  \caption{AES depth profiles of several spots along a combinatorial \ce{LaWN_{3-y}} film grown on a sapphire substrate. A) La cation content La/(W+La) and B) Oxygen anion content O/(O+N) as a function of sputtering time.  The inset in B) shows the position of the three spots measured by AES across the combinatorial sample (square symbols) as well as the two spots measured in synchrotron XRD (star symbols, see discussion below). The cation ratio for the XRD spots was measured with XRF.} 
  \label{fgr:AES}
\end{figure}

Cation composition La/(W+La) was mapped across the combinatorial samples using X-ray fluorescence (XRF), and several points were additionally measured with Auger electron spectroscopy (AES) depth profiling.  Consistently higher La/(W+La) values were measured via XRF in samples grown on transparent substrates (i.e., sapphire) compared to on pSi substrates, despite similar optical and electrical properties; to correct this, the AES measurements were calibrated with the XRF results from the films grown on pSi substrates.  Calibrated AES atomic \% depth profiles are shown in Figure S1. The overall calibrated composition derived from AES for the sample with mean La/(W+La) $=49.0(4)$\% on a sapphire substrate is \ce{La_{0.98}W_{1.02}N_{2.33}O_{0.07}} (Figure S1B), calculated assuming that the cation stoichiometry should sum to two. Similarly, a sample on a Si substrate with mean La/(W+La) $=50.6(1)$\% has overall composition \ce{La_{1.01}W_{0.99}N_{2.52}O_{0.07}} (Figure S1E). The observed nitrogen sub-stoichiometry is consistent with previous reports \cite{Talley2021}, and may be due to N loss during AES depth profiling.

Cation stoichiometry La/(W+La) and oxygen anion content O/(O+N) AES depth profiles of several spots along a combinatorial film grown on a sapphire substrate are shown in Figure \ref{fgr:AES}, and analogous depth profiles along a film grown on a pSi substrate are in Figure S2. The positions along the sample library at which AES depth profiles were measured are shown as the square symbols in the inset to Figure \ref{fgr:AES}B. The La content La/(W+La) is relatively flat through the film for all spots measured (Figure \ref{fgr:AES}A). The oxygen content, represented in Figure \ref{fgr:AES}B as O/(O+N), is low for all samples measured, below $\sim$7\%. The oxygen spectra are shown in Figure S3. For the La-rich sample with mean La/(W+La) $=52.5(4)$\%, the oxygen content is highest near the surface and lowest near the substrate interface, consistent with slight oxidation due to air exposure post-growth. For the W-rich sample with mean La/(W+La) $=48.8(7)$\%, the oxygen content rises slightly near the substrate interface, to $\sim$10\%. This may be due to either to pinholes allowing air to enter the interface between the sapphire substrate and the \ce{LaWN_{3-y}} film, or to small amounts of oxygen diffusing into the film from the sapphire substrate.  Optical SEM images of spots along combinatorial films grown on both sapphire and pSi substrates after annealing are shown in Figure S4.

\begin{figure}
\includegraphics[width=0.5\textwidth]{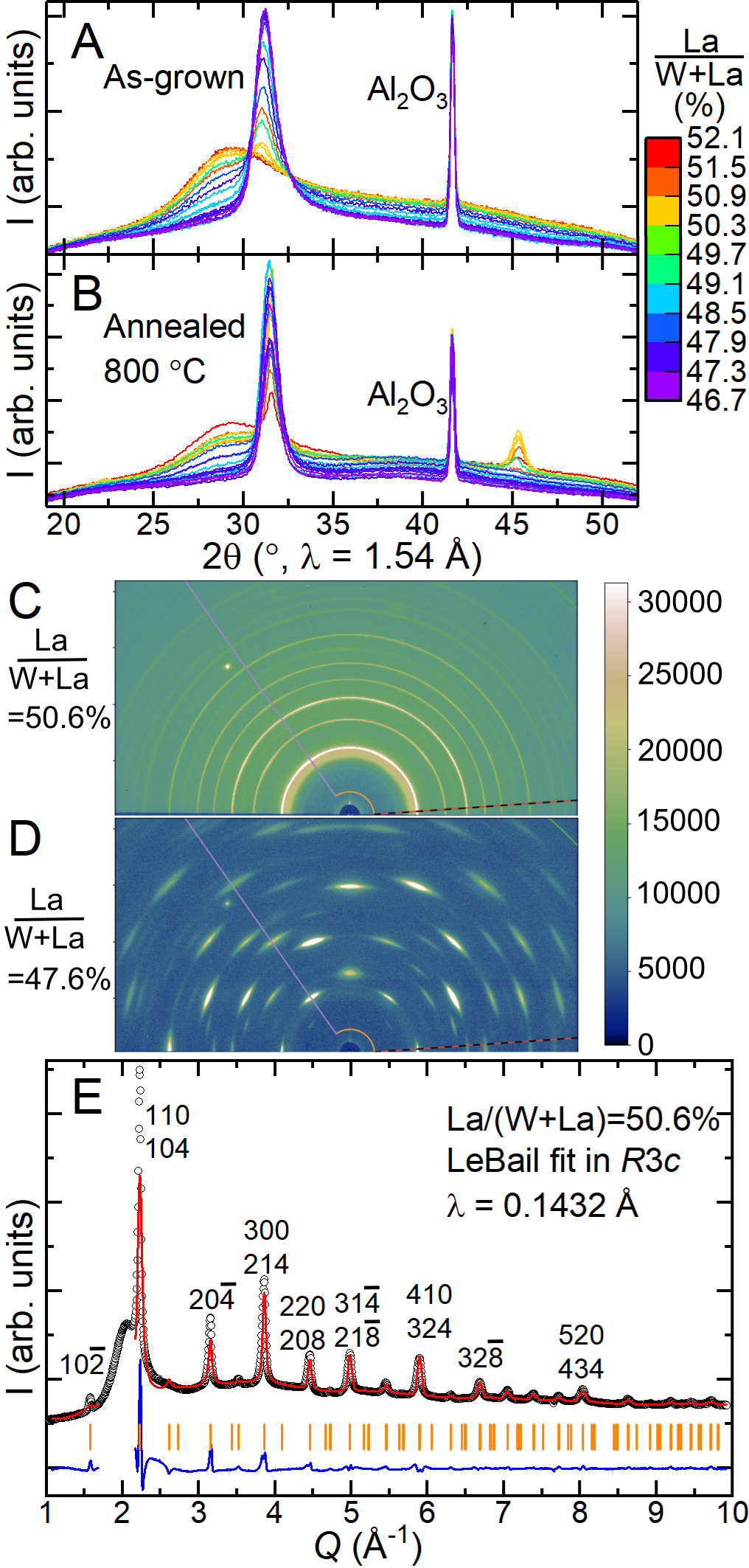}\\
  \caption{A,B) Laboratory XRD of a combinatorial \ce{LaWN_{3-y}} film grown on a sapphire substrate as a function of La/(W+La): A) as-grown, and B) after annealing at 800 $^\circ$C in flowing \ce{N2}. Calculated reflections for \ce{LaWN_3} are shown in B). The sharp peak  at  $2\theta =41.6^\circ$ is from the sapphire substrate. C,D) 2D detector images of grazing incidence XRD of C) La-rich (La/(W+La) = 50.6\%) and D) W-rich (La/(W+La) = 47.6\%) \ce{LaWN_{3-y}} combinatorial films collected at APS beamline 11-ID-B with wavelength $\lambda = 0.1432$ \AA. The films were grown on sapphire substrates, and the lines show the integration limits. The locations along the combinatorial sample at which these synchrotron XRD patterns were measured are shown in the inset of Figure \ref{fgr:AES}B. 
  E) LeBail fit in space group $R$3$c$ of the integrated synchrotron XRD data collected on the sample with La/(W+La) = 50.6\% shown in panel C). } 
  \label{fgr:XRD}
\end{figure}

We investigated whether \ce{LaWN_{3-y}} could be synthesized without an activated nitrogen plasma created with a nitrogen plasma source, which was used in the first report of thin film \ce{LaWN3} \cite{Talley2021} but is not standard equipment. We found that using a RF substrate bias, which controls the voltage of the substrate relative to the plasma, at a power of 50 W during deposition aided in forming nanocrystalline films that could then be crystallized via ex-situ annealing, as shown in Figure \ref{fgr:XRD}. In addition, RF substrate bias is a significantly more common capability compared to a nitrogen plasma source. The nanocrystalline films produced by this method contrast with the polycrystalline films observed in Ref. \cite{Talley2021} as a result of high-temperature growth with an activated nitrogen plasma in the same sputtering chamber; we attribute the difference either to a slight decrease in the maximum temperature achievable by the substrate heater or to the lower chemical potential of nitrogen with the RF substrate bias. To compensate for this, we performed ex-situ anneals in flowing \ce{N2} at 800 $^\circ$C for 10 min; this crystallized the perovskite phase, as shown in Figure \ref{fgr:XRD}B. The success of RF substrate biasing for \ce{LaWN_{3-y}} is consistent with previous results in other materials families; it has been found to increase reactive gas content in various films \cite{Brett1985}, affect the film stress \cite{Castellano1977}, and in particular for many sputtered nitride materials (including TiN and AlN) it has improved the electrical properties and film morphology \cite{Igasaki1980}.

\subsection{Crystal Structure}

Following the first synthesis step (deposition), we performed laboratory X-ray diffraction (XRD) on the combinatorial \ce{LaWN_{3-y}} films, as shown in Figure \ref{fgr:XRD}A. An intense but relatively broad peak at approximately $2\theta \approx 31^\circ$ ($Q=2.22$ \AA$^{-1}$) is observed in most of these as-grown XRD patterns; this is indicative of nanocrystalline perovskite formation. The average FWHM of this peak is 1.35(4)$^\circ~ 2\theta$, extracted from fits of the XRD data to a pseudo-Voigt function (see Table S1).  Samples with La/(W+La) $\gtrsim49\%$ La content also display a broad hump centered around $2\theta \approx 29^\circ$ ($Q=2$ \AA$^{-1}$), which corresponds roughly to the main diffraction peak of \ce{La2O3}, likely indicating the presence of nanocrystalline \ce{LaN} or \ce{La2O3}. The most La-rich (La/(W+La) $\gtrsim51\%$) samples display only this broad peak without a perovskite peak. The laboratory 2D detector data (Figure S6A) show that this broad peak in La-rich samples is relatively polycrystalline, displaying little to no crystallographic texture. The perovskite-related peak in W-rich samples, visible on the edges of the detector images, is textured; this is accompanied by an increase in the FWHM (see Figures S6 and S7).

Upon ex-situ annealing in flowing \ce{N2} at 800 $^\circ$C, additional perovskite peaks appear; this is shown for laboratory XRD in Figure \ref{fgr:XRD}B and for synchrotron grazing incidence XRD measured at beamline 11-ID-B at the APS in Figure \ref{fgr:XRD}C--E. The locations along the sample library at which synchrotron XRD was measured are shown in the inset of Figure \ref{fgr:AES}B. As a result of this crystallization, laboratory XRD (Figure \ref{fgr:XRD}B) reveals that the main perovskite peak---corresponding to the (110) and (104) reflections---``snaps'' to a more constant position of 31.501(2)$^\circ~ 2\theta$ and a narrower FWHM of 0.915(7)$^\circ~ 2\theta$. The positions and FWHM are shown in Figure S7, and the average values are in Table S1. This change is particularly evident in the La-rich samples; see Figures \ref{fgr:XRD}A--B and S5 as well as the 2D detector images in Figure S6A--B. All samples across the full range of cation stoichiometry La/(W+La) present in this combinatorial sample now display this main perovskite peak, although the trend from polycrystallinity in the La-rich samples to texture in the W-rich samples is still observed. 
 
Similar behavior as a result of annealing is observed in samples grown on pSi substrates, as shown in Figures S5 and S6C--D. No significant differences were observed between \ce{LaWN_{3-y}} grown on sapphire and pSi substrates in the peak position or FWHM in $2\theta$ ($Q$). However, in the radial direction ($\chi$), the average FWHM of the perovskite peak is narrower for films grown on sapphire (8.2(1)$^\circ$) compared to pSi (10.5(1)$^\circ$); this is visible in Figure S6.

A LeBail fit was performed on the integrated synchrotron data for the polycrystalline sample with La/(W+La) = 50.6\% in space group $R$3$c$, as shown in Figure \ref{fgr:XRD}E. The refined lattice parameters are $a=5.637(6)$ and $c=13.82(1)$, similar to the lattice parameters found in Refs. \cite{Talley2021} and \cite{MATSUISHI2022}. Some amorphous or nanocrystalline \ce{La2O3} is still present in the La-rich samples, as seen by the broad peak at $2\theta \approx 29^\circ$ in Figure \ref{fgr:XRD}B and $Q\approx2$ \AA$^{-1}$ in Figure \ref{fgr:XRD}E, but the overall amount appears lower compared to the as-grown samples, suggesting that at least some of it has been incorporated into the perovskite structure. However, as some La is tied up in the \ce{La2O3} impurity amorphous phase, the actual perovskite phase in the ``La-rich” samples may be La-poor. An extremely broad background centered at $2\theta \approx 40^\circ$ that emerges in the W-rich annealed samples additionally suggests that a small amount of nanocrystalline metallic W may be present. We note that polycrystalline \ce{LaWN_{3-y}} films synthesized previously by a two-step process (involving room-temperature growth with an activated nitrogen plasma followed by an ex-situ anneal at 900 $^\circ$C in flowing \ce{N2}) showed significant amounts of crystalline metallic W and small amounts of crystalline La and possibly WN \cite{Talley2021}. However, previous one-step growth at high-temperature with an activated nitrogen plasma yielded phase-pure \ce{LaWN_{3-y}} films, albeit with preferred orientation \cite{Talley2021}. Therefore, the crystallinity of these impurity phases is heavily dependent on synthesis method. The two-step technique employed here yields a range of crystalline textures for the \ce{LaWN_{3-y}} phase dependent on stoichiometry, but the impurity phases are reliably amorphous or nanocrystalline. 

\subsection{Optical Properties}
\begin{figure}
\includegraphics[width=0.95\textwidth]{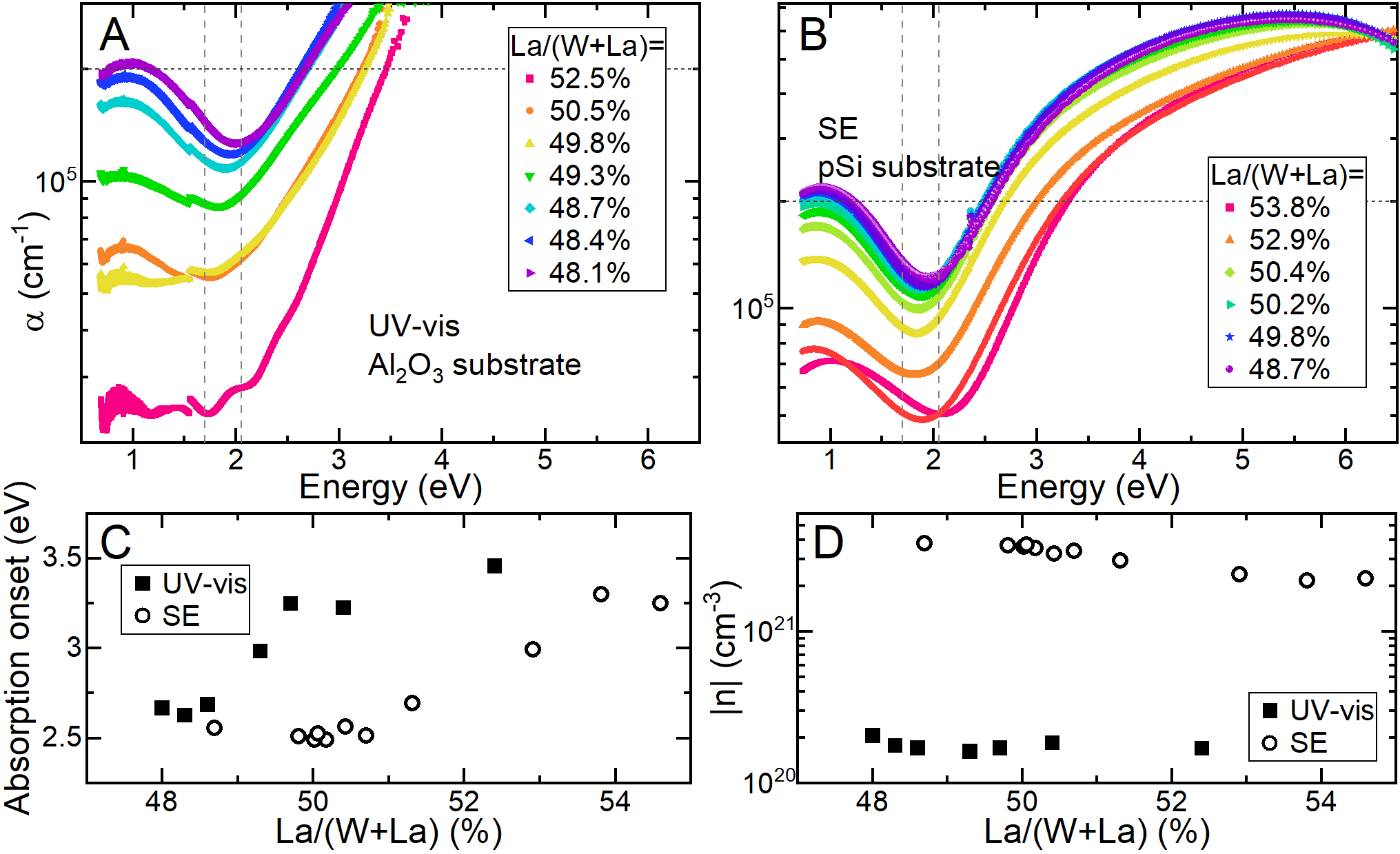}\\
  \caption{Absorption coefficient $\alpha$ extracted from A) UV-vis data of a combinatorial \ce{LaWN_{3-y}} film grown on a sapphire substrate, and B) spectroscopic ellipsometry (SE) data of a combinatorial \ce{LaWN_{3-y}} film grown on a pSi substrate. The horizontal dashed line is the chosen optical absorption onset. The vertical dashed lines represent the calculated band gaps at 1.7 eV (indirect) and 2.05 eV (optical). C) Absorption onsets extracted from UV-vis and SE data of \ce{LaWN_{3-y}} grown on sapphire and pSi substrates, respectively. The absorption onset is calculated as the energy at which each curve intersects with $\alpha = 2\times10^5$ cm$^{-1}$ (see dashed lines in panels A and B). D) Carrier densities $|n|$ extracted from the low-energy regions of the SE and UV-vis data.  } 
  \label{fgr:Optical}
\end{figure}

To probe the optical properties of these films, we performed UV-vis measurements across a combinatorial \ce{LaWN_{3-y}} film grown on a sapphire substrate and spectroscopic ellipsometry (SE) measurements across a film grown on a pSi substrate. The absorption coefficients ($\alpha$) extracted from both measurements are shown in Figure \ref{fgr:Optical}A and B, respectively. Modeling the SE data revealed that a few hundred nm of \ce{LaWN_{3-y}} does not transmit photons. The two techniques display similar trends overall, with the La-rich compositions exhibiting absorption onsets at higher energies than the W-rich compositions.

To quantify the trends with composition yet avoid the errors endemic to Tauc analyses, we define an optical absorption onset as the energy when $\alpha = 2\times10^5$, shown in Figure \ref{fgr:Optical}C. While some variation exists between the onsets extracted from the UV-vis and SE data, both techniques yield low onsets for La/(W+La) $< 50$\%, at approximately 2.5 eV. The absorption onset is highly sensitive to composition. It increases with higher La/(W+La) values, plateauing at about 3.3 -- 3.5 eV. The behavior at La/(W+La) $<50$\% is likely due to microscopic metallic W or \ce{WN_x} impurities in the W-rich region, while the behavior in the La-rich region is likely dominated by insulating \ce{La2O3} impurities.  This is accompanied by a color change from translucent yellow-brown to black (as shown in Ref. \cite{Talley2021}). We used the GW approximation to calculate an indirect band gap of 1.70 eV and an optical (vertical transition) band gap = 2.05 eV (see Methods for details), which is in the range of band gaps (1.59 -- 2.0 eV) obtained from DFT calculations using hybrid functionals\cite{Sarmiento-Perez2015,Fang2017,Singh2018,MATSUISHI2022,Geng2023}. However, these values are slightly different than those reported for the bulk, high-pressure sample (1.2 eV direct and 2.2 eV indirect), perhaps reflecting different defects. \cite{MATSUISHI2022}. Due to the choice of $\alpha = 2\times10^5$ as a cutoff, the extracted absorption onsets are higher than the calculated band gap, but the trends should be consistent.

Significant sub-gap absorption is evident for W-rich samples in the 1--2 eV region in both the UV-vis and SE data (Figure \ref{fgr:Optical}A and B). In the SE data, this region was modeled with a Drude oscillator, allowing the extraction of carrier densities ($|n|$), as shown in Figure \ref{fgr:Optical}D. These are on the order of $|n|\approx$ 2 -- 4$\times10^{21}$ cm$^{-3}$, which is high but consistent with the large observed absorption below the gap. The densities are higher for W-rich samples. The peak in the low-energy UV-vis absorption can be interpreted as the plasma frequency ($\omega_p$); carrier densities ($|n|$) were extracted as:
\begin{equation}\label{plasma}
    |n| = \frac{\omega_p^2 m^* \epsilon_\infty \epsilon_0} {e^2}
\end{equation}
where $m^*$ is the electron effective mass and $\epsilon_\infty$ is the static dielectric constant. Here, we use DFT-calculated $m^*$ = 0.836 $m_e$ and $\epsilon_\infty$ = 13.61 (see Methods for details). These values are similar to those calculated with DFT methods \cite{Liu2020,Geng2023}. This calculation yields $|n|$ values on the order of $|n|\approx2\times10^{20}$ cm$^{-3}$, slightly lower than the values extracted from the SE fitting and exhibiting fewer trends with composition. 

\subsection{Electrical Transport}
\begin{figure}
\includegraphics[width=0.95\textwidth]{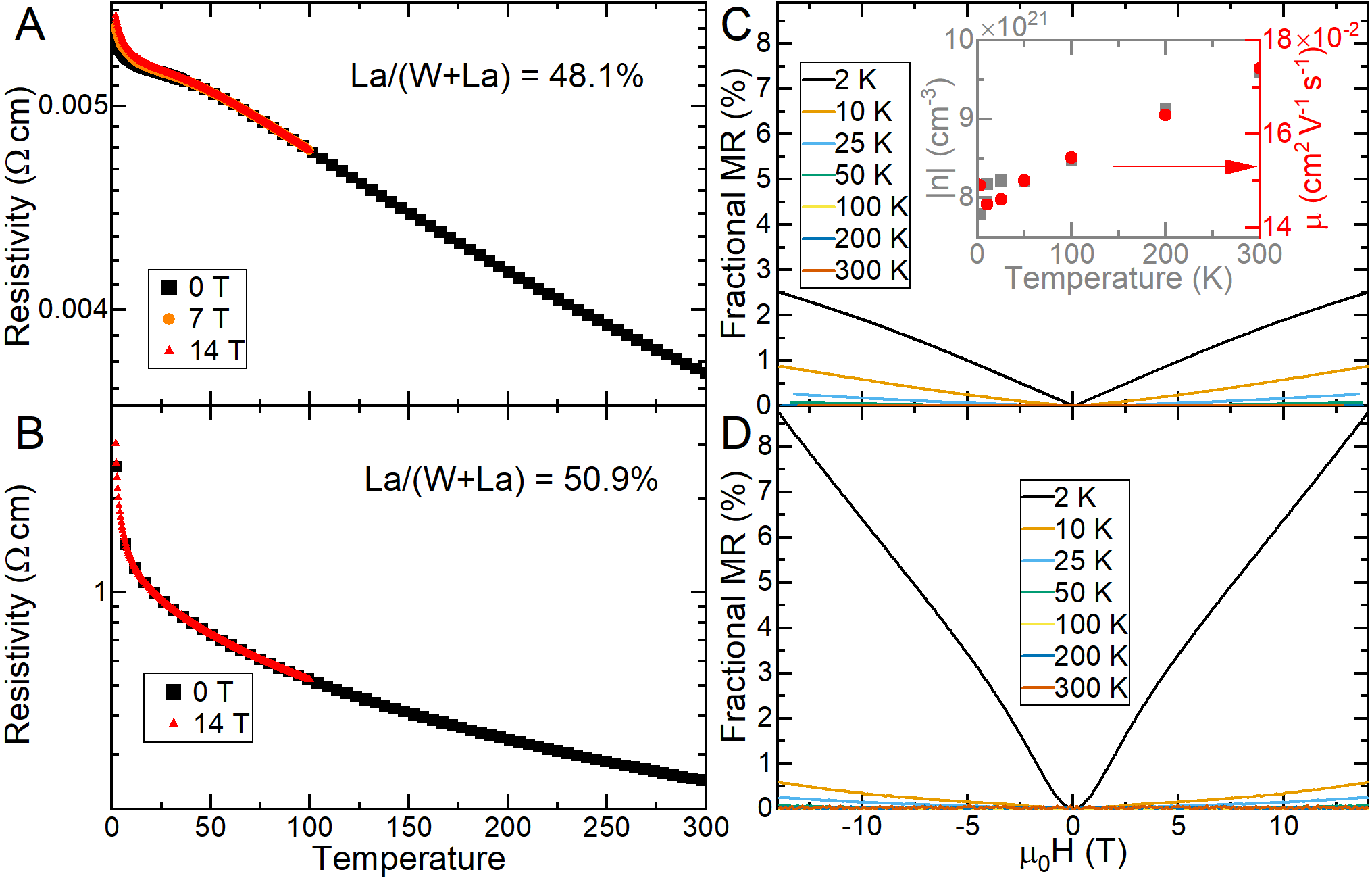}\\
  \caption{Temperature-dependent resistivity ($R_S$) of \ce{LaWN_{3-y}} films with A) La/(W+La) $=48.1$\% and B) measured under several applied fields.  
  C) Fractional magnetoresistance (MR) of \ce{LaWN_{3-y}} films with C) La/(W+La) $=48.1$\% and D) La/(W+La) $=50.9$\% measured at several temperatures. The inset in C) shows the extracted carrier densities ($n$) and mobilities ($\mu$) as a function of temperature.} 
  \label{fgr:PPMS}
\end{figure}

We measure the temperature and magnetic field-dependent electrical transport of \ce{LaWN_{3-y}} films for the first time. 
Figure \ref{fgr:PPMS} summarizes the electrical transport measured on two \ce{LaWN_{3-y}} samples grown on sapphire substrates, a W-rich sample (La/(W+La) $=48.1$\%, Figure \ref{fgr:PPMS}A,C) and a La-rich sample (La/(W+La) $=50.9$\%, Figure \ref{fgr:PPMS}B,D). Strikingly, the measured resistivities ($\rho$) of the two samples---which have a composition difference of only La/(W+La) $\approx3$\%---differ by approximately two orders of magnitude.
The W-rich sample exhibits temperature-dependent resistivity consistent with degenerately doped semiconducting behavior (Figure \ref{fgr:PPMS}A). The La-rich sample is approximately two orders of magnitude more insulating and exhibits a temperature dependence characteristic of an insulator. This difference in resistivity is large given the small difference in cation ratio between the two samples, less than 3\% La/(W+La).  While the samples measured here do contain significant levels of defects, no clear superconducting transition is seen below $T\approx 6$ K in either sample, implying low amounts of---or very amorphous---\ce{WN_x} impurities, in contrast to the reported bulk \ce{LaReN3} sample that contained superconducting \ce{ReN_x} impurities\cite{Klos2021}.

The behavior below $T\approx 50$ K in the W-rich sample (Figure \ref{fgr:PPMS}A) may be due to freezing out hopping between conductive regions of impurities such as amorphous or nanocrystalline W or \ce{WN_x}.  This low-temperature region was measured in several applied fields; the curves deviate from the $\mu_0H = 0$ T data below approximately $T=50$ K. Following prior work\cite{Kabilova_2019,Culman2022}, we investigated several models to fit the low-temperature, zero field region (below $T=46$ K), including a mix of defect-mediated contributions in the form of Mott 3D variable range hopping (VRH) and band-mediated contributions in the form of an Arrhenius model: $\rho(T) = \rho_0 + Ae^{\frac{\Delta E}{k_BT}} + Be^{(\frac{T_0}{T})^{1/4}}$. However, the Arrhenius contribution did not improve the fit, so we fit the data with only the 3D VRH model (see Figure S8). We hypothesize that at low temperatures, the conduction is dominated by defect states, and as the temperature decreases the thermal energy becomes competitive with the energy of hopping.  The temperature trends in the extracted carrier densities ($n$) and mobilities ($\mu$), discussed below, are consistent with this hypothesis (see inset of Figure \ref{fgr:PPMS}C). 

With only a small change in cation ratio ($<$3\%), the La-rich sample (La/(W+La) = 50.9\%) is more insulating than the W-rich sample by approximately two orders of magnitude (Figure \ref{fgr:PPMS}B). The low-temperature behavior is characteristic of an insulator, consistent with the likely impurities of \ce{La2O3} or \ce{LaN} rather than conductive W or \ce{WN_x}. We note that the presence of \ce{La2O3} impurities implies that the perovskite phase in this nominally La-rich sample may be La-poor, implying a potential source of unintentional electrons from N vacancies and O substitution. We also attempted fitting the low-temperature behavior ($T<38$ K) with a mixed Arrhenius and 3D VRH model. Similar to the W-rich sample, the Arrhenius component did not fit the data well (see Figure S8). The extracted 3D VRH coefficients ($B$) from fitting these two samples differ by four orders of magnitude ($\sim$2.9E-4 for the W-rich sample and $\sim$2 for the La-rich sample), reflecting the difference in their resistivities.

Magnetic field-dependent resistance was measured at a range of temperatures, shown as the fractional magnetoresistance (FMR = $[R(\mu_0H) - R(0)]/R(0) \times 100$) in Figure \ref{fgr:PPMS}C--D. At $T=2$ K, the FMR of the W-rich sample (Figure \ref{fgr:PPMS}C) is linear and reaches approximately 2.5\% at $\mu_0H = 14$ T. The origins of this linear MR could be semi-classical or impurity/defect-driven\cite{Parish2003,Hu2008,Song2015,Kisslinger2017}. The linear MR decreases quickly at higher temperatures and disappears by $T=50$ K, consistent with the temperature-dependent results (Figure \ref{fgr:PPMS}A). By measuring the Hall resistance ($R_{xy}$, Figure S9), we find that the W-rich film is n-type with a carrier density of $n = -7.8\times10^{21}$ cm$^{-3}$ and a low carrier mobility of $\mu = 0.149$ cm$^2$ V$^{-1}$ s$^{-1}$) at $T=2$ K. The trends with temperature are shown in the inset of Figure \ref{fgr:PPMS}C.  This n-type behavior likely arises from nitrogen vacancies and oxygen defects, and it is consistent with observed behavior in many nitrides \cite{Bauers2019,Rom2023}. The extracted $|n|$ values are comparable to the sheet carrier densities extracted from the optical measurements (Figure \ref{fgr:Optical}), although there is better agreement with the SE values than with the UV-vis values. The FMR of the La-rich sample (Figure \ref{fgr:PPMS}D) is linear and larger than the W-rich sample, $\sim$9\% at $T=2$ K. 

\subsection{Photoresponse}
\begin{figure}
\includegraphics[width=0.5\textwidth]{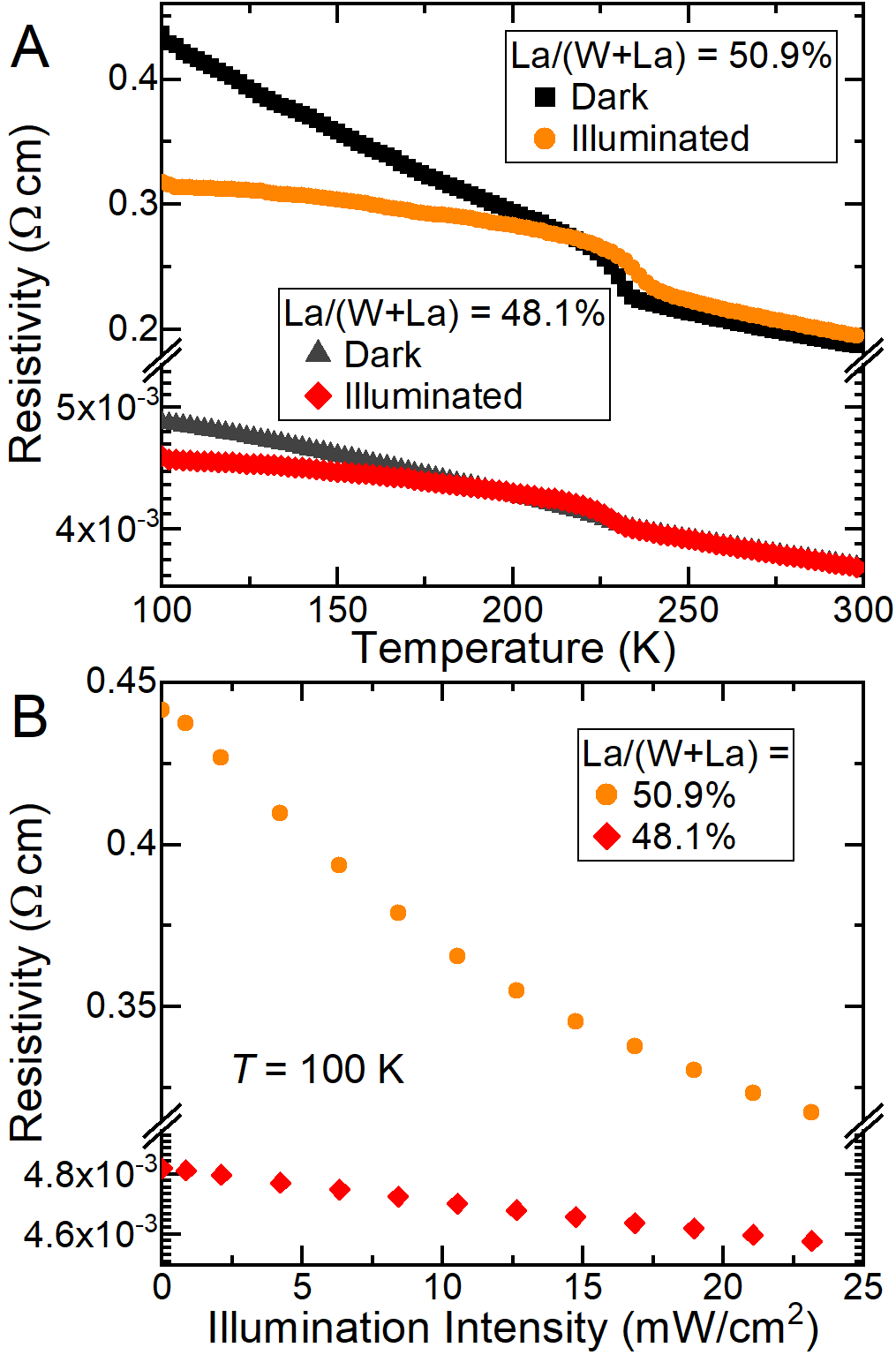}\\
  \caption{A) Temperature-dependent resistivity of two samples of \ce{LaWN_{3-y}} with La/(W+La) $=48.1$\% and La/(W+La) $=50.9$\% in the dark and under AM1.5G spectrum illumination at an intensity of 23 mW/cm$^2$. B) Resistivity as a function of illumination intensity of the AM1.5G spectrum. } 
  \label{fgr:Electronic}
\end{figure}

As we calculated a bandgap for \ce{LaWN3} of 1.7 eV (indirect) using the GW approximation, we measured the temperature-dependent resistivity of these two samples of \ce{LaWN_{3-y}} under AM1.5G spectrum illumination in a different chamber. The light that reached the sample location was calibrated with a known CIGS solar cell, revealing that the highest effective illumination was approximately 23 mW/cm$^2$. The dark resistivities of both samples between $T=100$ -- 300 K, shown in Figure \ref{fgr:Electronic}A, are consistent with the PPMS measurements. The change in slope at $T\approx$ 225 K is an instrumental artefact; it has been observed across a range of samples on several substrates\cite{Bauers2019,Bauers2020,Rom2023}.  

Under illumination, the resistivities of both samples decrease. This effect is most pronounced at lower temperatures, below approximately $T=175$ K. We examined the dependence of resistivity upon illumination intensity at $T=100$ K, as shown in Figure \ref{fgr:Electronic}B, finding a smooth evolution up to the highest possible intensity. For the W-rich sample, the change ($[\rho(I) - \rho(0)]/\rho(0) \times 100$) is approximately 5\% at the highest illumination, while for the La-rich sample the change is $\sim$28\%. We note that the switching dynamics of this photoresponse are slow, suggesting that the underlying behavior is not carrier recombination but more likely trapping and de-trapping of defect states at the surface of the film. This hypothesis is consistent with the observed stronger response at low temperatures: the energy of trapping must reach rough equivalence with thermal energy at approximately 150 -- 175 K, where the dark and illuminated resistivities merge. The different defects---and defect densities---in the W-rich sample (metallic W and \ce{WN_x}) compared to the La-rich sample (insulating \ce{La2O3}) likely contribute to the different degree of photoresponse (5\% and 28\%, respectively). These results are broadly consistent with recent computational work arguing that nitride perovskites are defect-intolerant semiconductors \cite{Geng2023}. 
This is the first demonstration of photoresponse in \ce{LaWN_{3-y}} films, and further investigation will be required to fully understand the nature of the carrier dynamics and effects of defects in \ce{LaWN_{3-y}} films. 

\subsection{Magnetism}

\begin{figure}
\includegraphics[width=0.5\textwidth]{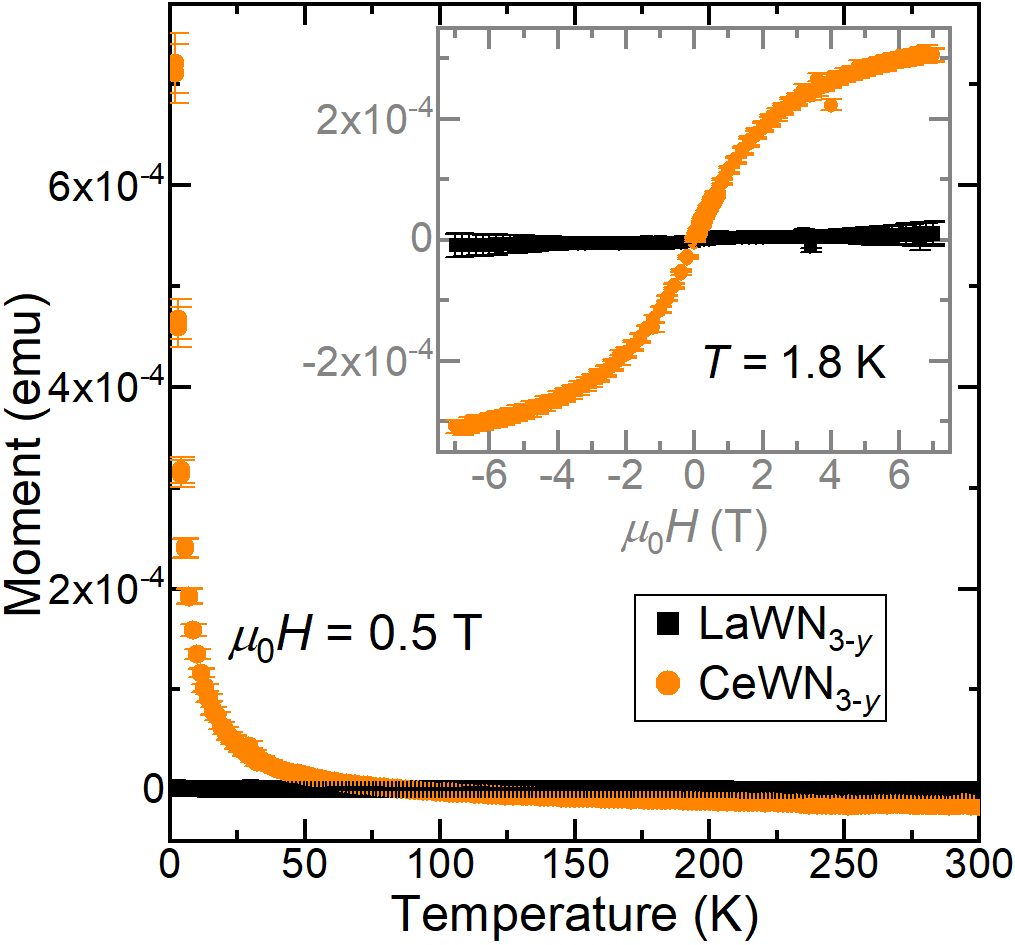}\\
  \caption{Comparison of DC magnetic moment measurements at an applied magnetic field of $\mu_0H = 0.5$~T for \ce{LaWN_{3-y}} and \ce{CeWN_{3}}.  Inset: Moment as a function of applied field at $T=1.8$ K for \ce{LaWN_{3-y}} and \ce{CeWN_{3}}. The substrate contribution has been subtracted from all data, and the resulting signal for \ce{CeWN_{3}} was scaled to match the approximate volume of the \ce{LaWN_{3-y}} film.  The measured films have cation stoichiometries La/(W+La) $=50.6$\% and Ce/(W+Ce) $=51$\%, as measured by XRF, and were grown on a pSi substrate with W on the back (\ce{LaWN_{3-y}}) and a pSi substrate (\ce{CeWN_{3-y}}).} 
  \label{fgr:magneticSusceptibility}
\end{figure}

To check the magnetism of \ce{LaWN_{3-y}}, we performed temperature-dependent DC moment measurements on a piece of a combinatorial \ce{LaWN_{3-y}} film grown on a pSi substrate at an applied magnetic field of $\mu_0H = 0.5$ T, as shown in Figure \ref{fgr:magneticSusceptibility}. The cation stoichiometry of this sample is approximately La/(W+La) $=50.6$\%, as measured by XRF. We compare it to the Ce analog, \ce{CeWN_{3-y}}, which we recently reported as a new nitride perovskite that is paramagnetic down to $T=2$ K\cite{Sherbondy2022}. Although the mass of these perovskite thin films cannot be quantified precisely, to yield the most quantitative comparison the substrate contribution for each sample (pSi with W on the back for \ce{LaWN_{3-y}} and \ce{SiN_x}|pSi for \ce{CeWN_{3-y}}) has been subtracted, and the resulting perovskite signal has been scaled to the same approximate film volume. We note, however, that the moment values are still approximate, and that both films may have some amorphous or nanocrystalline phase fraction in addition to the crystalline perovskite phase.

The measured moment for \ce{LaWN_{3-y}} is temperature-independent and very weakly positive at $\mu_0H = 0.5$ T, while the \ce{CeWN_{3-y}} sample shows a much stronger paramagnetic response. The inset shows the field-dependent measurement at $T=1.8$ K for both perovskites; similarly, the \ce{LaWN_{3-y}} response is very weak, while the \ce{CeWN_{3-y}} sample is paramagnetic. The slope of the \ce{LaWN_{3-y}} signal is nearly flat within the error bars, with a small ferromagnetic contribution at low fields that likely arises from sample holders and handling. A temperature-dependent scan was performed in a low field of $\mu_0H = 0.0002$ T to check for a drop in resistance due to superconducting impurities, such as the \ce{ReN_x} impurities observed in the bulk \ce{LaReN3} synthesized via high-pressure techniques\cite{Klos2021}; as shown in Figure S10, none are observed down to $T=1.8$ K. 

For ideal \ce{LaWN_{3}}, we expect diamagnetism  consistent with \ce{La^{3+}} and \ce{W^{6+}}. Interestingly, although \ce{LaReN3} should be paramagnetic due to \ce{Re^{6+}}, it was reported to exhibit weak Pauli paramagnetism, consistent with its predicted and reported metallicity\cite{Flores-Livas2019,Klos2021}. The magnetic behavior observed here for \ce{LaWN_{3}} is a very weak response---especially compared with \ce{CeWN_{3-y}}, which exhibits paramagnetism consistent with the Curie law---that could be consistent with either diamagnetism or possibly Pauli paramagnetism from impurity phases. Comparing \ce{LaWN_{3}} with \ce{LaReN3}, \ce{CeWN_{3-y}} and \ce{CeMoN_{3-y}}, which has a transition to long-range antiferromagnetism below $T_N\approx8$ K\cite{Sherbondy2022}, begins to illuminate broader trends in the magnetism of nitride perovskites. For the Ce perovskites, we were not able to experimentally determine whether the magnetism arises from \ce{Ce^{4+}} or \ce{W^{5+}}/\ce{Mo^{5+}}, but the differences between them as well as the weak behavior of both La perovskites suggest that the magnetic moment in nitride perovskites arises primarily from the $A$ site. 

\section{\label{sec:Conclusion}Conclusion}
We have investigated the structural, optical, and electronic properties of combinatorial thin films of \ce{LaWN_{3-y}}, one of the first oxygen-free nitride perovskites ever reported. As previous work on thin film \ce{LaWN_{3-y}} focused on its experimentally-proven piezoelectricity but encountered problems measuring a ferroelectric response due to large leakage current, a detailed investigation into its optoelectronic properties is necessary to better understand and subsequently control the underlying physics of \ce{LaWN_{3-y}}. We grow combinatorial thin films on both silicon and sapphire substrates using a two-step process: a high-temperature deposition employing a RF substrate bias, yielding amorphous or nanocrystalline films. An ex-situ anneal at 800 $^\circ$C in flowing \ce{N2} crystallizes the perovskite phase. Texturing is dependent on stoichiometry, with polycrystalline La-rich films and highly textured W-rich films. AES depth profiles demonstrate a low amount of oxygen through the film, confirming the viability of this synthesis method for producing fully nitrided \ce{LaWN_{3-y}}.

We find that the optical and electronic properties of thin film \ce{LaWN_{3-y}} are highly sensitive to stoichiometry. UV-vis spectroscopy and spectroscopic ellipsometry reveal that the absorption onset, which should display similar trends to the bandgap, changes by $\sim$1 eV. In W-rich samples (La/(W+La) $<50$\%), significant sub-gap absorption is observed, consistent with metallic W or \ce{WN_x} impurities. We report the low-temperature resistivity of thin film \ce{LaWN_{3-y}} in applied magnetic fields from 0 -- 14 T, revealing behavior consistent with a degenerately doped semiconductor in slightly W-rich \ce{LaWN_{3-y}} and insulating behavior in slightly La-rich \ce{LaWN_{3-y}}. The fractional magnetoresistance is linear and small, consistent with defect scattering, and a W-rich sample exhibits n-type carriers with high densities $n\approx-8\times10^{21}$ cm$^{-2}$ and low mobilities $\mu\approx$ 0.149 cm$^2$/(V s) at $T=2$ K. A photoresponse is observed: the resistivity decreases at low temperatures under solar spectrum illumination, and a much larger effect is observed for the La-rich sample compared to the W-rich sample; this is consistent with a defect trapping mechanism. Magnetic moment measurements of \ce{LaWN_{3-y}} reveal a very weak magnetic response, consistent with expectations.

The sensitivity of the measured structural and optoelectronic properties of \ce{LaWN_{3-y}} to cation stoichiometry suggests that it is a line compound, which is common in oxide perovskites. This has important implications for the successful functionalization of the physical properties of \ce{LaWN3}, including its high predicted ferroelectric response: precise composition control and thorough characterization will be crucial. For example, to reduce leakage for ferroelectric applications, polycrystalline films with a higher band gap---to avoid shunts through grain boundaries---are generally preferable. In light of these results, future work on ferroelectric \ce{LaWN3} should focus on optimizing La-rich compositions, which we show prefer to grow polycrystalline. They may also be more likely to form with insulating phases at grain boundaries, affording low leakage, although improving materials quality and engineering defects will also be necessary. Overall, the semiconducting physical properties and sensitive composition dependence of \ce{LaWN3} investigated here are likely generalizable to other members of this emerging materials family of nitride perovskites as a whole, which has been the focus of many computational studies but only a few experimental reports. Thus, we provide a path forward to understanding the intrinsic physics of this new materials family and functionalizing its predicted properties.

\begin{acknowledgements}
This work was authored by the National Renewable Energy Laboratory (NREL), operated by Alliance for Sustainable Energy, LLC, for the U.S. Department of Energy (DOE) under Contract No. DE-AC36-08GO28308. This work was supported by the Laboratory Directed Research and Development (LDRD) Program at NREL. Funding supporting development and operation of synthesis and characterization equipment was provided by the DOE Office of Science, Office of Basic Energy Sciences. Funding supporting GW calculations was provided by the DOE Office of Science collaboratively from the Office of Basic Energy Sciences, Division of Materials Science and the Advanced Scientific Computing Research (ASCR) program. This work used computational resources sponsored by the Department of Energy’s Office of Energy Efficiency and Renewable Energy, located at NREL. Magnetometry was performed through NSF DMR 1905909. This research used resources of the Advanced Photon Source, a U.S. Department of Energy (DOE) Office of Science user facility operated for the DOE Office of Science by Argonne National Laboratory under Contract No. DE-AC02-06CH11357. The authors thank G.L. Brennecka for helpful discussions; R. Sherbondy for growing the \ce{CeWN_{3-y}} film; O.L. Borkiewicz for assistance at APS beamline 11-ID-B; and C.L. Rom for assistance with sample masking.
The views expressed in the article do not necessarily represent the views of the DOE or the U.S. Government. 
\end{acknowledgements}

R.W.S., K.R.T., and A.Z. conceptualized the work; R.W.S. performed thin film growth and characterization (XRD, XRF), with assistance and input from J.S.M. (ellipsometry), I.L. (resistivity and magnetoresistance), M.H. (UV-vis), and J.C. and C.P.M. (photoresistivity). C.L.P. performed the AES measurements and analysis.  S.E. provided access to the MPMS3. S.R.B. and A.Z. advised project directions. R.W.S. wrote the manuscript with input from all authors.

%


\end{document}